\documentclass[trackchanges,twocolumn]{aastex7}

\usepackage{amssymb}
\usepackage{amsmath}	
\usepackage{mathtools}
\usepackage{gensymb}
\usepackage{color}
\usepackage{enumitem}
\usepackage{textcomp}
\usepackage{multirow}
\usepackage{adjustbox}
\usepackage{booktabs}
\usepackage{color}
\usepackage{epsfig}
\usepackage{color}
\usepackage{graphicx}
\usepackage{longtable}
\usepackage{hyperref}
\usepackage{soul}

\begin{document}

\title{Different polarized components of the quasar 3C 286 revealed by FAST}

\author[gname=pengfu,sname=Tian]{Pengfu Tian}
\affiliation{Department of Physics, Guizhou Minzu University, Guiyang 550025, China;}
\affiliation{Department of Astronomy, School of Physics and Technology, Wuhan University, Wuhan 430072, China;}
\email[show]{tpengfu@gzmu.edu.cn}  

\author[gname=Xiao, sname=Chen]{Xiao Chen} 
\affiliation{Department of Physics, College of Science, Southern University of Science and Technology, Shenzhen 518055, China;}
\affiliation{Department of Astronomy, School of Physics and Technology, Wuhan University, Wuhan 430072, China;}
\email{chenxiao@sustech.edu.cn}

\author[,gname=Wen, sname=Yang]{Wen Yang} 
\affiliation{Department of Astronomy, School of Physics and Technology, Wuhan University, Wuhan 430072, China;}
\email{yangwen63@whu.edu.cn}

\author[gname=Wei, sname=Wang]{Wei Wang} 
\affiliation{Department of Astronomy, School of Physics and Technology, Wuhan University, Wuhan 430072, China;}
\email[show]{wangwei2017@whu.edu.cn}
\correspondingauthor{Pengfu Tian, Wei Wang}

\begin{abstract}

3C 286, a well-known radio calibrator, exhibits the stability in both of total flux density (FD) and polarization parameters. However, its stable and luminous interstellar radio signal may encounter interplanetary scintillation (IPS) due to density irregularities in the solar wind within the heliosphere. In this work, we analyze high-time-resolution observations of 3C 286 obtained with the Five-hundred-meter Aperture Spherical radio Telescope (FAST) from 2019 to 2023. Our analysis reveals that IPS affects the polarized flux densities of the Stokes I, Q, and U parameters, whereas Stokes V shows no detectable IPS-induced variations. The IPS variations detected in Stokes I are synchronous with those in Stokes U, while those in Stokes Q exhibit greater randomness. The crosscorrelation function (CCF) results indicate no time delay between Stokes I and U, but a delay of approximately 2.8 seconds between Stokes I and Q. This suggests that the different polarized radio emissions of 3C 286 originate from distinct emission regions, specifically the core and the southwestern jet. Furthermore, the projections of the radio core and jet component onto the scintillation screen at 1 AU yield a solar wind plasma speed of $\sim 637$ km/s.


\end{abstract}



\section{Introduction} 
Quasars as a subclass of active galactic nuclei (AGN) with extremely activity and luminosity in multiband radiation \citep{1963Natur.197.1040S,1964ApJ...140..796S}, are generally much brighter than the host galaxy in the optical band \citep{2019arXiv191104305F}. 3C 286 is a well-known compact steep spectrum (CSS, $< 15$ kpc) quasar with a $z=0.849$ redshift \citep{1985A&A...143..292F,1995A&AS..112..235A} and frequently used as a radio calibrator, exhibiting stable total flux densities with 8-9$\%$ linear polarization (LP), negligible circular polarization (CP), and a linear polarization angle (PA) of approximately 33 degrees at the L band \citep{1994A&A...284..331O,2013ApJS..206...16P}. According to the polarization and intensity images observed by the Very Large Array (VLA) \citep{1995A&AS..112..235A,2004ChJAA...4...28A,2013ApJS..204...19P}, this quasar occupies about 3.8 arcsec at 1.66 GHz in the east-west direction \citep{1989MNRAS.240..657S}, exhibits a non-linear triple structure which can be roughly resolved into the core, eastern and southwestern three major components \citep{1989ApJ...347..713H,1995A&AS..112..235A}. The southwestern and east jet components located approximately 2.6 arcsec and 0.8 arcsec from the core respectively. The different components show different polarized behaviors. The core exhibits a strong dominance of Stokes U \citep{1996A&A...312..380J}, while the Stokes Q is primarily contributed by the ejected jets \citep{1995A&AS..112..235A}. Comparatively, the southwestern jet is brighter than the eastern jet and shows a higher linear polarization percentage, suggesting that the Stokes Q predominantly originates from the southwestern jet \citep{2016ApJ...824..132N}.

When observing quasars such as the compact radio source 3C 286 at small angular distances from the Sun in the sky, the influence of the solar wind must be accounted for. Typically, radio waves emitted by compact radio objects with angular size $\lesssim$ 1 arcsec undergo scattering by small-scale (10–1000 km) electron-density fluctuations within the solar wind plasma in the heliosphere \citep{2015SoPh..290.2539M}. This scattering manifests as flux density intensity variations of the radio source when observed by ground-based radio telescopes—a phenomenon known as interplanetary scintillation \citep[IPS]{1964Natur.203.1214H}.

Normally, The propagation of solar winds with the turbulent density will lead to random fluctuations in time and frequency domains of radio signals from compact radio sources, which is important for understanding space weather and coronal mass ejections \citep{2023AdSpR..72.5341J,2023AdSpR..72.5361C}. Moreover, IPS observations of radio sources with a small diameter can also offer insights into the properties of solar winds, like speed or density \citep{1978RaSc...13..591C,1967ApJ...147..449C,2008JGRA..113.6307T,2015SoPh..290.2539M,2016AdSpR..57.1307C,2021MNRAS.504.5437L,2022ApJS..259....2P}. In addition, the IPS signals may also contain the information of different emission regions of the compact sources. The previous FAST observations resolved that the source 3C 286 exhibited time-domain convolution of emissions from multiple compact components \citep{2021MNRAS.504.5437L}. 

Based on the high sensitivity of polarization observational ability for FAST, we will study the different polarized components from 3C 286 in this work. This paper is structured as follows. In Section 2, we present the observational data utilized in this study, along with the detailed data reduction and analysis  methodologies employed. The relevant results would be presented in Section 3. Subsequently, we discuss our findings including the calibration effect and solar winds in Section 4. Finally, in Section 5, we summarize our results.

\section{Observations and Data Reduction}\label{sec:reduction}

\subsection{Observations}
FAST, the world's largest single-dish radio telescope, provides a system sensitivity of 2000 m$^2$/K and high time resolution, which are essential for detecting faint and rapid radio signals. For this observation, we used its 19-beam pulsar backend receiver, which covers a frequency range of 1–1.5 GHz with 4096 channels and records all four Stokes parameters \citep{2020RAA....20...64J}. At a representative observing frequency of 1.25 GHz and a declination of $30^\circ$, the full width at half power of the FAST beam is 2.95 arcmin.

Our observations were carried out with high power mode noise injected every 983.04 microseconds to ensure the calibration precise, while the public observation data are collected in Tracking mode. The injected noise, however, is a reference signal with certain polarization properties (100$\%$ linear polarized and 0$\%$ circular polarized), can be useful for the calibration of correction of Muller matrix \citep{2001PASP..113.1243H,2023Natur.621..271T,2023JHEAp..39...43T}. 

\begin{table*}[h]
    \centering
    \caption{
    FAST observations of 3C 286 from 2019 to 2023 with the elongation angle $\epsilon$ between 3C 286 and the Sun about $35^\circ-70^\circ$. FD and POL indicate the flux and polarization calibration can be carried out for that observation. St., Res. and Du. represent the start time (UTC), resolution and duration respectively.} 
    \label{tab:observations}
    \renewcommand\arraystretch{1.5}
    {\begin{tabular}{c | c c c c c c c c c}
    \hline
    PID & Date & St. & MJD & Res. ($\mu$s) & Du. (s) & Obs. mode & Calibration & IPS & $\epsilon$ ($^\circ$)\\
    \hline
    \multirow{3}{*}{3059}
    & 2019-12-05 & 01:02:21 & 58822 & 49.152 & 180 & Tracking & \textemdash, \textemdash & \textemdash & 75.57\\
    & 2019-12-10 & 01:01:28 & 58827 & 49.152 & 180 & Tracking & FD, POL & Yes & 75.34\\
    & 2019-12-10 & 23:50:20 & 58827 & 49.152 & 180 & Tracking & FD, POL & Yes & 75.30\\
    \hline
    \multirow{5}{*}{PT2020\_0123}
    & 2020-09-26 & 05:59:37 & 59118 & 196.608 & 2400 & Tracking & FD, \textemdash & Yes & 62.29\\
    & 2020-09-27 & 06:26:31 & 59119 & 196.608 & 3600 & Tracking & FD, \textemdash & Yes & 62.25\\
    & 2020-09-28 & 05:50:23 & 59120 & 196.608 & 2400 & Tracking & FD, \textemdash & Yes & 62.21\\
    & 2020-10-16 & 03:22:23 & 59138 & 196.608 & 1800 & Tracking & FD, \textemdash & Yes & 61.44\\
    & 2020-10-30 & 03:25:35 & 59152 & 196.608 & 1800 & Tracking & FD, \textemdash & Yes & 60.84\\
    \hline
    \multirow{1}{*}{PT2020\_0154}
    & 2021-07-15 & 10:58:16 & 59410 & 196.608 & 2400 & Tracking & \textemdash, \textemdash & Yes & 50.45\\
    \hline
    \multirow{1}{*}{DDT2021\_7}
    & 2022-06-16 & 12:22:58 & 59746 & 98.304 & 150 & OnOff & FD, POL & Yes & 39.80\\
    \hline
    \multirow{1}{*}{PT2022\_0134}
    & 2022-09-07 & 04:44:09 & 59829 & 196.608 & 1800 & Tracking & \textemdash, \textemdash & Yes & 37.99\\
    \hline
    \multirow{2}{*}{PT2022\_0202}
    & 2022-09-30 & 06:50:10 & 59852 & 98.304 & 150 & OnOff & FD, POL & Yes & 37.56\\
    & 2022-10-21 & 02:54:10 & 59873 & 98.304 & 150 & OnOff & FD, POL & Yes & 37.21\\
    \hline
    \multirow{3}{*}{PT2023\_0086}
    & 2023-09-06 & 05:29:10 & 60193 & 98.304 & 150 & OnOff & FD, POL & Yes & 35.78\\
    & 2023-09-12 & 05:42:35 & 60199 & 98.304 & 150 & OnOff & FD, POL & Yes & 35.83\\
    & 2023-11-17 & 00:30:10 & 60265 & 98.304 & 150 & OnOff & FD, POL & Yes & 36.54\\
    \hline
    \end{tabular}}
\end{table*}

Including several publicly available datasets, our observations of 3C 286 span from 2019 to 2023. All information pertaining to observations is listed in Table \ref{tab:observations}. The observations in PID of 3059, PT2020\_0123, PT2020\_0154 and PT2022\_0134 are public data of FAST, monitored with Tracking mode. 
The observations conducted in PT2020\_0154 and PT2022\_0134 lacked observations of sky background and standard noise injection. Consequently, the calibration of flux and polarization was unattainable. Observations in PT2020\_0123 have only two polarization channels AA and BB, which means this observation can be merely calibrated in flux density. Note, the initial observation in PID 3059 on the 5th of December in the year 2019, rendering it non-calibratable, is predominantly attributable to erroneous noise injection, wherein the noise signal was injected solely into the BB polarization channel. 

\subsection{RFI elimination} 

\begin{figure}[htbp]
    \centering
    \includegraphics[width=0.45\textwidth]{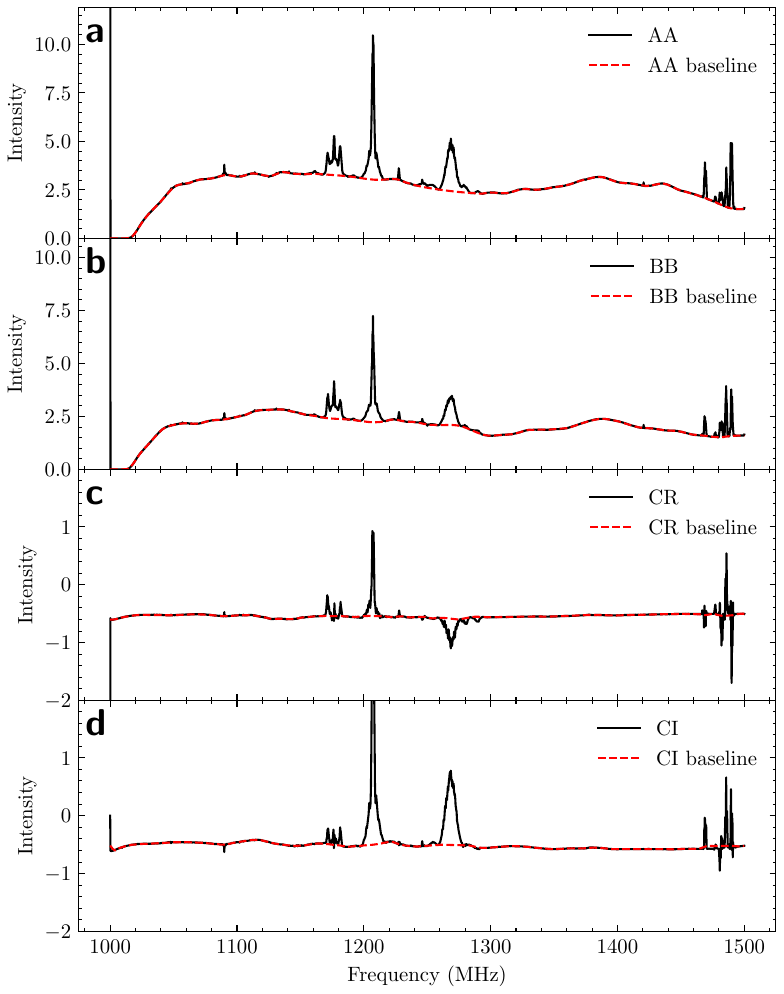}
    \caption{The original temporally averaged bandpass spectra (black solid lines) and corresponding baselines (red dashed lines) fitted by ArPLS algorithm for AA ({\bf a}), BB ({\bf b}), CR ({\bf c}), and CI ({\bf d}).}
    \label{fig:bpArPLS}
\end{figure}

RFI elimination is always the most important task in radio observation data reduction as it would contaminate the observation data seriously. Fortunately, 3C 286 is bright in the L band, making it less affected by RFIs compared to other weak radio sources. Therefore, the RFI elimination of 3C 286 in our observations only requires the direct removal of some RFI channels, which can be easily identified from the spectrum, as can be seen in Figure \ref{fig:bpArPLS} around 1200 MHz.

However, the complicated RFI environment makes the reduction of offsource observation data difficult. In this work, we utilized the Asymmetrically reweighted Penalized Least Squares smoothing algorithm \citep[ArPLS,]{baek2015baseline} to model the baseline of the spectrum in offsource observations. This approach is necessary for obtaining stable $OFF$, $OFFCAL$ and further $T_{sys}$ values according to equation \ref{eq:Tsys}.

In Fig. \ref{fig:bpArPLS}, we present the baselines of time-averaged offsource bandpass spectra of AA, BB, CR, and CI (original fits data format), estimated by the ArPLS algorithm. Consequently, we can use the fitted baselines to replace the spectra of offsource observations for further $T_{sys}$ calculation. It is worth noting that RFIs would be eliminated during the subtraction of $OFFCAL - OFF$, because the noise diode remains either continuously on or alternative on and off states, then the RFI environment will be the same for $OFF$ and $OFFCAL$. However, as RFI also varies with time and frequency, even if the subtraction can remove RFI, the numerator $OFF$ in equation \ref{eq:Tsys} is still RFI-polluted. Hence, direct baseline fitting to obtain stable $T_{sys}$ improves the efficiency and accuracy of calibration processing. Another aspect concentrates on the baseline fitting of the CR and CI spectra. The ArPLS algorithm works by adjusting the weights assigned to RFI iteratively to obtain the final result, this approach is contradictory for RFI lower than the baseline \citep{wang2024cpl}. To solve this problem, we developed a method in which RFI occurring below the baseline is inverted above it, thereby facilitating its recognition by the ArPLS algorithm. The baseline identification performs well, as shown in Figure \ref{fig:bpArPLS}.

The methods discussed above has a good performance for computing the baseline of off-source spectral. Furthermore, owing to the concurrent recording of data from all 19 beams of the 19-beam receiver during observation projects, the majority of RFIs can be identified and effectively mitigated through careful scrutiny of the data derived from the non-central beams. 

\subsection{Flux calibration}

There are several observations with massive data and two observation modes. Regardless of changes in observation modes or different parameters, the fundamental method of calibration remains unchanged. Firstly, it is necessary to calculate the relevant telescope temperatures.

$T_{sys}$ is the temperature of telescope when it is directed towards the sky background, which can be written as:
\begin{equation}
\begin{split}
    T_{sys}=T_{cal}\cdot\frac{\overline{OFF}}{\overline{OFFCAL}-\overline{OFF}}
\end{split}
\label{eq:Tsys}
\end{equation}
where $T_{cal}$ is the temperature caused by the injected noise, which in our observations is approximately 12.5 K for high power mode. $\overline{OFFCAL}$ and $\overline{OFF}$ are the averaged machine counts integrated over the period when the noise injection diode switched on and off as the telescope is staring at the sky background.

Moreover, the temperature of telescope when it is observing the target source $T_{onsource}$ is,
\begin{equation}
\begin{split}
    T_{onsource} (t)=T_{cal}\cdot\frac{ON (t)}{\overline{ONCAL}-\overline{ON}}
\end{split}
\label{eq:Tonsource}
\end{equation}
where $\overline{ONCAL}$ and $\overline{ON}$ are the averaged machine counts integrated over the period when the noise injection diode switched on and off as the telescope is staring at the target source. In practice, this interval used for mean values estimation can be selected manually. Consequently, the temperature variations $T_{src}$ induced by observing the calibrator is,
\begin{equation}
\begin{split}
    T_{src} (t)=T_{onsource} (t)-T_{sys}
\end{split}
\end{equation}

However, some observations are designed with no noise injected when the telescope is observing the target source, in which $T_{src}$ can be further estimated by \citep{2004hpa..book.....L},
\begin{equation}
\begin{split}
    T_{src} (t)=T_{sys}\cdot\frac{ON (t)-\overline{OFF}}{\overline{OFF}}
\end{split}
\end{equation}

Once the temperature $T_{src}$ induced by the observing source is computed, the next step is to calculate the telescope gain to determine the true flux density. Nevertheless, this is somewhat difficult for calibrator 3C 286. Typically, radio calibration involve primary computation of the telescope temperature variations induced by observing the calibrator, followed by comparing these temperature variations with the true flux of the calibrator, thereby deriving a telescope gain $G$. Subsequently, based on this gain $G$, one can estimate the true flux density of other sources observed during the same observation session.


During our observations, we noted that the linear polarization of 3C 286 often displayed a slightly amplified magnitude compared to its actual measurement. Conversely, the polarization observed in the IPS-quiet region remained consistent with its true polarization. This phenomenon suggests that the full-bandwidth signal is polarized, and the segment devoid of IPS signal can be regarded as representative of the actual 3C 286 flux and polarization. Thus, by utilizing the 'low-IPS' data segment and comparing it with the actual flux of 3C 286, we obtained the telescope gain G, 
\begin{equation}
\label{eq:FD0}
\begin{split}
    G=\frac{T_{src}^\prime}{FD_0}
\end{split}
\end{equation}
where the $T_{src}^\prime$ is the $T_{src}$ estimated from 'low-IPS' data segments, which are selected as those data points in the dynamic spectrum whose flux density falls below the threshold of $\mu -\sigma$ ($\mu$ and $\sigma$  represent the mean value and standard deviation of the total flux density). $FD_0$ is the true total flux density of 3C 286.

Gain $G$ was subsequently applied to the $T_{src}$ of that observation to derive the flux density. In practice, calibration can also be achieved by observing another calibrator. However, in previous observations, the presence of 3C 286 eliminated the additional observations of calibrators throughout the calibration process.

The calibration method employed above inevitably introduces errors when calculating the absolute flux density of the source unless another calibration source is observed. Just like the increases caused by selecting 'low-IPS' regions where $\sigma$ grows as observations approach the Sun, the final flux must therefore be corrected by a factor of 1/($\mu-\sigma$). Additionally, the steep spectrum of 3C 286 means that varying good frequency channel selections significantly impact the final flux density. For instance, RFI contamination limited the observation on MJD 60193 to 1284–1450 MHz, resulting in a lower measured flux density.


\subsection{Polarization calibration} 
Typically, the characterization of the polarization state of a signal is accomplished through the comprehensive representation provided by the Stokes vector. However, in practical scenarios, disparities may arise between this vector and the intrinsic Stokes vector due to instrumental response and polarization leakage caused by the observing system, and can be characterized by the Mueller matrix $\mathcal{M}$. The Mueller is combination of a set of independent matrices which are coming from different part of observation process, and can be treated as a transformation between the measured and intrinsic Stokes vectors\citep{2025wang-pol},
\begin{equation}
\begin{split}
    S_{m}=\mathcal{M}\times S_{i}=
    \begin{bmatrix}
    1 & f & 0 & 0\\
    f & 1 & 0 & 0\\
    0 & 0 & cos\Delta\psi & -sin\Delta\psi \\
    0 & 0 & sin\Delta\psi & cos\Delta\psi
    \end{bmatrix}\times S_{i}
\end{split}
\end{equation}
where the $S_{m}$ and $S_{i}$ are the measured and intrinsic Stokes vectors respectively. $f$ and $\Delta\psi$ are differential gain and phase, which can be estimated from the injected 100$\%$ linear polarized noise signal.

The polarization parameters are then computed from the intrinsic Stokes vector $S_{i}$, 
\begin{equation}
\begin{split}
    LP & = \frac{\sqrt{Q^2+U^2}}{I}\\
    CP & = \frac{V}{I}\\
    PA & = 0.5\times tan^{-1}(\frac{U}{Q})\\
\end{split}
\end{equation}

It is worth noting that the composition of the Mueller matrix typically involves several independent matrices, such as $\mathcal{M}_{Amp}$, $\mathcal{M}_{CC}$, $\mathcal{M}_{Feed}$, and $\mathcal{M}_{PA}$. However, the 19-beam receiver is designed to rotate with the celestial sphere, rendering the P.A. (parallactic angle) term $\mathcal{M}_{PA}$ unnecessary. Moreover, since the feed of FAST is dual linear, the Feed term $\mathcal{M}_{Feed}$ equals 1. As for the cc (caused by the coupled orthogonal probes) term $\mathcal{M}_{cc}$, it is primarily responsible for the leakage from the total intensity to Stokes V. However, this is negligible for FAST, hence the cc term is also not a concern. In this study, the generation of the Mueller matrix is primarily attributed to the amplifier chains. Consequently, polarization calibration predominantly entails the correction of the Amp term $\mathcal{M}_{Amp}$ \citep{2023JHEAp..39...43T}.

The polarization characteristics of 3C 286 are stable, as evidenced by its LP values of 8.6$\%$ at 1.05 GHz and 9.5$\%$ at 1.45 GHz \citep{2013ApJS..204...19P}, which closely align with our measurements. Additionally, our results confirm the absence of circular polarization in 3C 286. However, a notable discrepancy arises in the PA of 3C 286 in some observations, typically registering at 33 degrees in the FAST band, whereas our observations frequently yield PAs from 40 to 50 degrees.  

\section{Results}\label{sec:res}
\begin{figure}[htbp]
    \centering
    \includegraphics[width=0.45\textwidth]{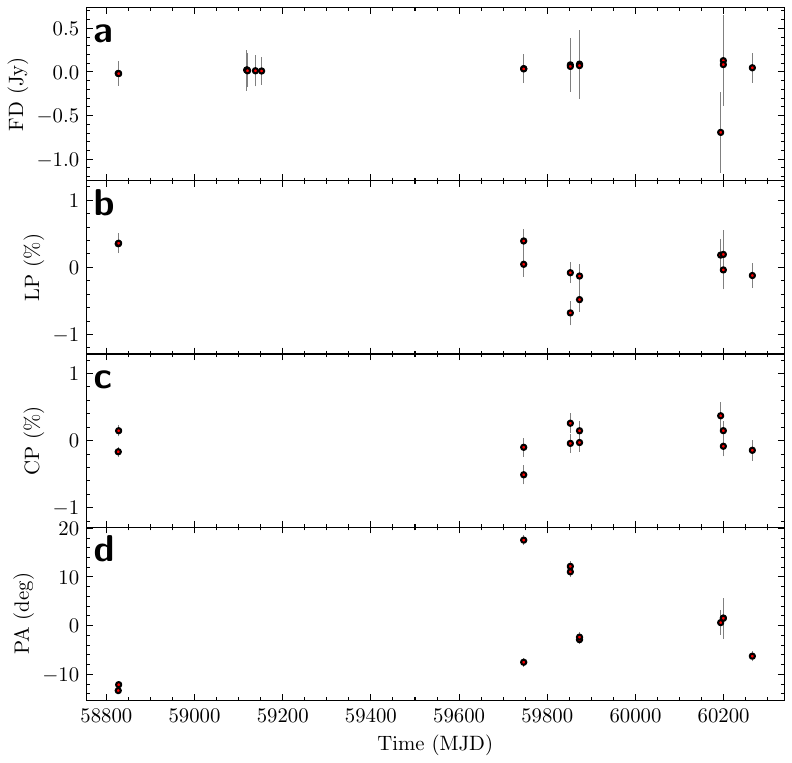}
    \caption{Long-term monitoring of quasar 3C 286 from 2019 to 2023 with FAST. Shown are the deviations of the total flux density (\textbf{a}), linear polarization (\textbf{b}), circular polarization (\textbf{c}), and linear polarization position angle (\textbf{d}) from their respective best-estimated mean values. The error bars indicate the fluctuations of time series on second timescales, which are attributed to IPS caused by solar wind turbulence.}
    \label{fig:LC}
\end{figure}

Based on the data reduction and calibration procedures described above, the calibrated results are shown in Figure \ref{fig:LC}. Quasar 3C 286, known for its stable flux density and compact structure, is adopted as a flux-density calibrator \citep{2013ApJS..204...19P}. Over the period from 2019 to 2023, the total flux density $I$ and fractional linear polarization $L$ remain stable. In Figure \ref{fig:LC}, we show the deviations of total flux density, linear polarization, circular polarization, and linear polarization position angle from their best-estimated mean values to highlight the measurement uncertainties. The error bars represent fluctuations induced by IPS. The larger scatter observed in PA compared to FD and LP arises from the uncorrelated IPS in Stokes Q and U.

\begin{figure}[htbp]
    \centering
    \includegraphics[width=0.45\textwidth]{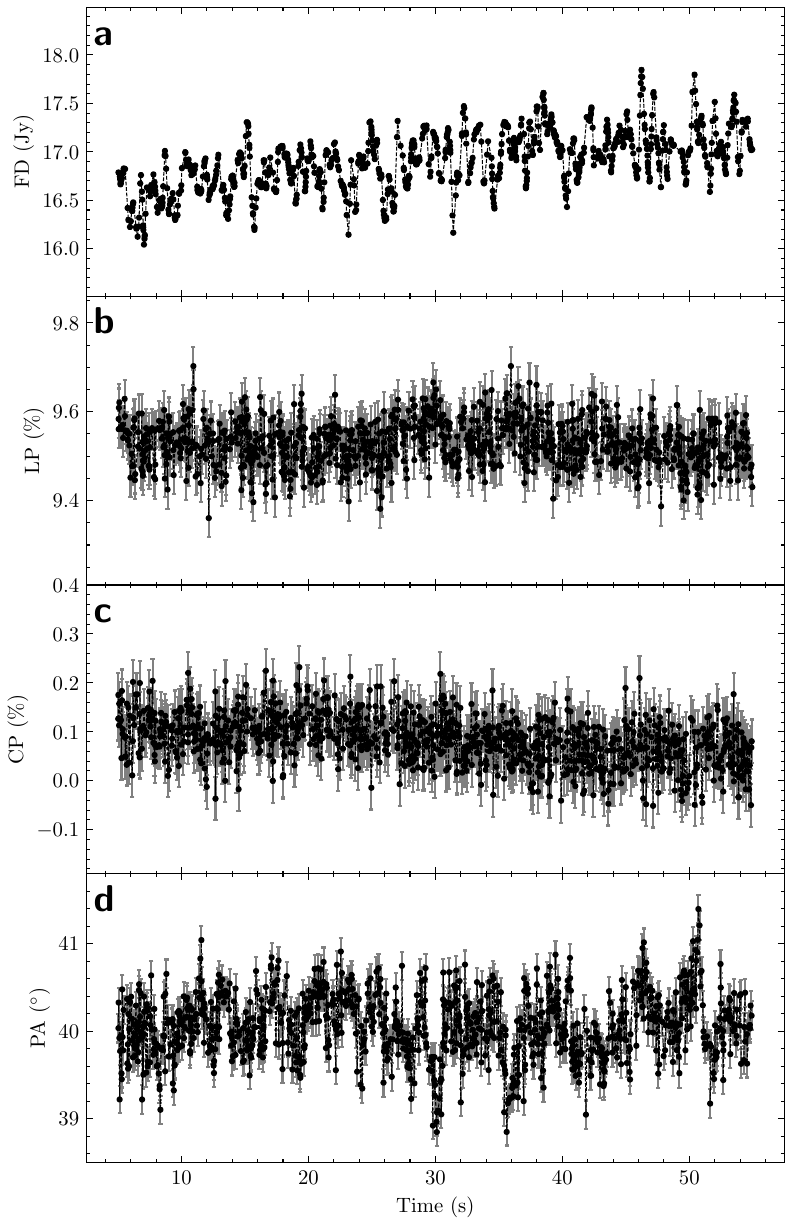}
    \caption{The light curves of total flux density ({\bf a}), linear polarization ({\bf b}), circular polarization ({\bf c}), and polarization position angle ({\bf d}) of 3C 286 from the FAST observations conducted on September 30, 2022. The comparison with Figure \ref{fig:LC} shows that while FD and LP remain stable, the liner polarization position angle exhibits larger apparent fluctuations due to uncorrelated IPS-induced variations in Stokes $Q$ and $U$. The scintillation index of FD in panel \textbf{a} is $m=0.018$.}
    \label{fig:lightcurves}
\end{figure}

Figure \ref{fig:lightcurves} presents a zoomed-in time series of the four Stokes parameters for the observation PT2022$\_$0202, confirming that the enhanced PA scatter is caused by IPS rather than intrinsic variability. The source shows a stable total flux density of $\sim16$–17 Jy, a linear polarization of $\sim 9\%$, negligible circular polarization, and a mean PA of $\sim40^\circ$.

\begin{figure}[hbp]
    \centering
    \includegraphics[width=0.45\textwidth]{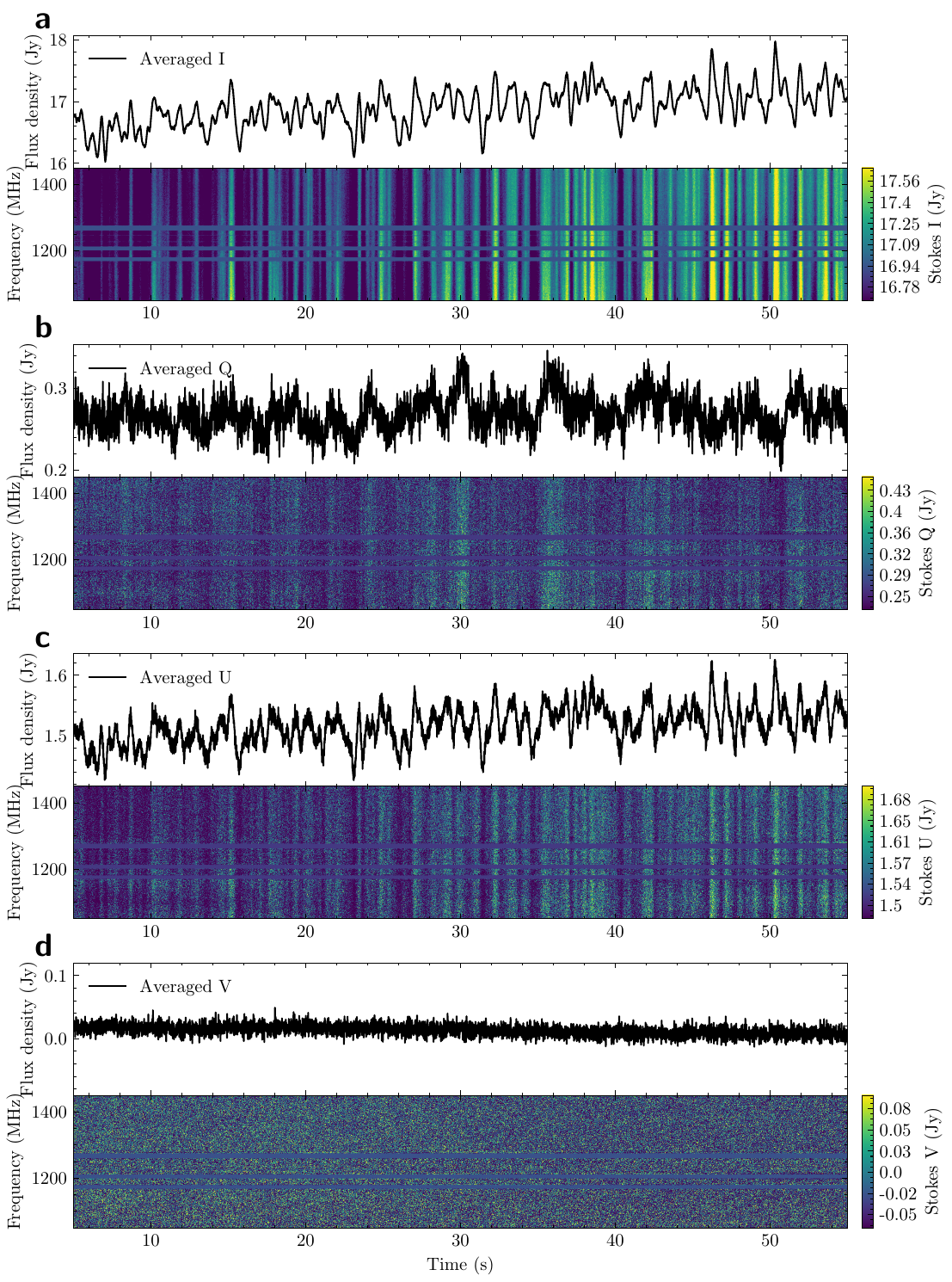}
    \caption{The averaged flux densities and dynamic spectra of Stokes I ({\bf a}), Q ({\bf b}), U ({\bf c}), and V ({\bf d}) from the FAST observations conducted on September 30, 2022. The obscured horizontal stripes indicate the RFIs mask. Stokes U shows variations coherent with Stokes I, while Stokes Q, although exhibiting some similar variations in some time intervals, primarily manifests as relatively stochastic fluctuations. The Stokes V is near zero for all observation.}
    \label{fig:Dynamic}
\end{figure}

\begin{figure*}[htp]
    \centering
    \includegraphics[width=0.32\textwidth]{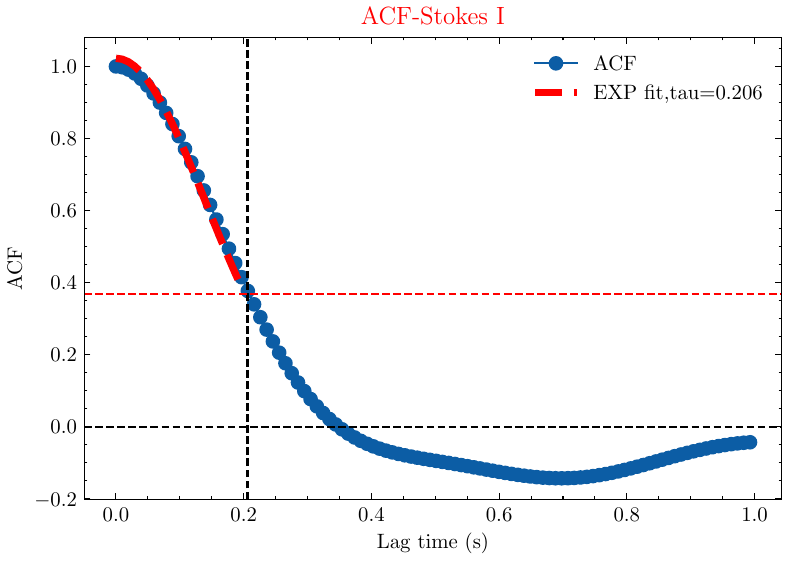}
    \includegraphics[width=0.32\textwidth]{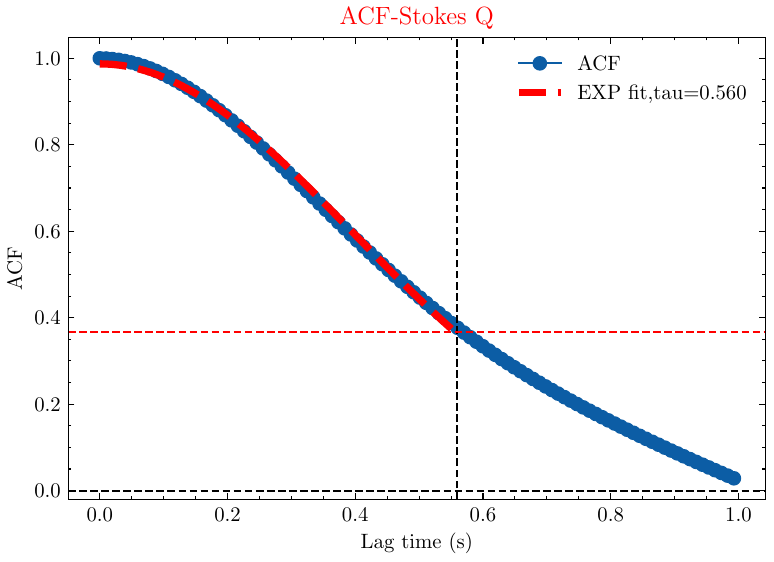}
    \includegraphics[width=0.32\textwidth]{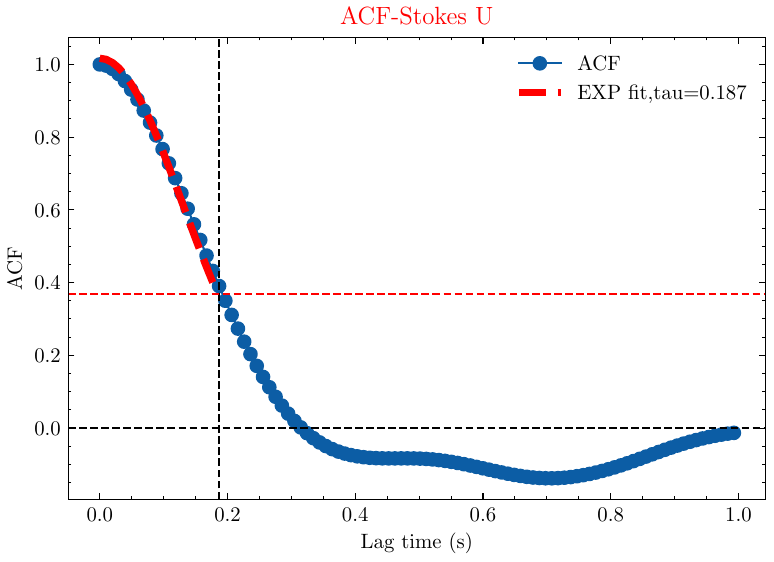}
    \caption{From left to right: the ACF results of Stokes I, Q and U. The bold red dashed curves represent Gaussian fits, and the horizontal red dashed lines mark the e-folding level. The derived characteristic timescales are $\tau_I$ = 0.206 s, $\tau_Q$ = 0.560 s, and $\tau_U$ = 0.187 s. The scintillation indices of Stokes I, Q, and U are $m_I = 0.018$, $m_Q = 0.079$, and $m_U = 0.020$, respectively.}
    \label{fig:acf}
\end{figure*}

On the other hand, with the full-channel polarization recording capability of FAST and the calibration method described in Section \ref{sec:reduction}, we obtained light curves and dynamic spectra for all four Stokes parameters. Using the observation PT2022$\_$0202 as an example, Figure \ref{fig:Dynamic} demonstrates that the IPS signals detected by FAST are broadband. And, some of these signals exhibit frequency drifting, which can be primarily attributed to the frequency dependence of the Fresnel scale ($r_{F}=\sqrt{\lambda D/2\pi}$, where D is the distance between scintillation screen and the Earth). The dynamic spectra reveal that the IPS patterns in Stokes I and U are nearly identical, while IPS in Stokes Q displays a distinctly different variability pattern. The flux density of Stokes Q (0.2–0.4 Jy) is likely underestimated due to the off-center positioning of its emitting regions relative to the telescope’s beam center. Meanwhile, the absence of IPS in Stokes V is consistent with 3C 286’s negligible circular polarization—a result logically expected from the underlying IPS mechanism.

\section{Discussion}\label{sec:dis}
Early VLA, Multi-Element Radio Linked Interferometer Network (MERLIN) and the Atacama Large Millimeter and Submillimeter Array (ALMA) observations established the kiloparsec-scale radio morphology of 3C 286 \citep{1989MNRAS.240..657S,1995A&AS..112..235A,2016ApJ...824..132N,2017ApJS..230....7P}. Moreover, this source is well-documented as a stable radio-band calibrator, exhibiting stability in both flux density and polarization properties across multiple emission components \citep{2004ChJAA...4...28A,2013ApJS..204...19P,2017ApJS..230....7P}. 3C 286 is also classified as a CSS object with a spectral index of approximately -0.61 \citep{2017MNRAS.466..952A}. Therefore, the frequency-averaged flux densities in Figure \ref{fig:LC} show fluctuations across different observation epochs because of varying RFI environments. 

The effect of IPS on the Stokes parameters is briefly summarized here. First, IPS is an intensity modulation and it is the same for linear and circular polarization. The Stokes parameters are sums or differences of intensities of various linear or circular polarization, so they are also modulated by IPS. Second, IPS is a spatial variation, which appears as a time variation as the spatial variation moves across the line of sight to the source. If there are two sources displaced by an angle $\theta$, there will be two diffraction patterns on the observing plane, a distance $z$ from the scattering region, displaced in distance by $\theta*z$, if the velocity is aligned with this displacement the scintillations will be the same but delayed. Otherwise the scintillations of two sources will be uncorrelated if the pattern displacement exceeds the spatial scale.

The angular distance between 3C 286 and the Sun also critically affects calibrator observations. As demonstrated by \citep{2010SoPh..265..309M,2013ARep...57..586G,2023AdSpR..72.5341J}, the scintillation index $m$ and power density spectrum, which characterize the degree of scintillation, decreases with increasing solar elongation and angular size of source. Our observations confirm this relationship. As shown in Figure \ref{fig:Dynamic}, the IPS dynamic spectra for Stokes parameters I and U exhibit finer temporal structure than that for Stokes Q. This is more likely due to the larger spatial angular size of the SW jet component, which dominate the emission in Stokes Q. Its more extended spatial structure leads to angular broadening, suppressing the amplitude of IPS fluctuations and smoothing finer scintillation patterns, rendering them undetectable in Stokes Q. In Figure \ref{fig:acf}, we quantified the IPS timescales by analyzing the autocorrelation functions (ACFs) of these light curves denoised by empirical mode decomposition \citep[EMD]{1998RSPSA.454..903H}. Fitting a Gaussian model to each ACF and measuring its e-folding time yielded the characteristic timescales: $\tau_I$ = 0.2 s, $\tau_Q$ = 0.56 s, and $\tau_U$ = 0.19 s. On the other hand, the normalized deviations for Stokes I, Q and U after noise removed are 0.018, 0.079, and 0.020 respectively, indicating a relatively higher degree of fluctuation in Stokes Q. This result is consistent with the larger spatial structure of the SW jet responsible for the emission in Stokes Q, which consequently leads to greater variability in its observed IPS signal.

The angular separation between the core and the southwestern jet is 2.6 arcsec \citep{1995A&AS..112..235A,2004ChJAA...4...28A}, corresponding to a projected distance of $\sim$ 1885 km on the scintillation screen at 1 AU, which significantly exceeds the Fresnel scale ($\sim$ 189 km) at this location. In this scenario, the core and southwestern jet can be treated as a discrete point sources (angular size $\lesssim$ 1 arcsec), with their radio emissions experience independent scattering though inhomogeneous solar wind plasma at distinct regions of scintillation screen. Consequently, their scintillation patterns may become correlated but temporally delayed as the same plasma irregularity successively traverses their respective LOS.

\begin{figure*}[htbp]
    \centering
    \includegraphics[width=0.45\textwidth]{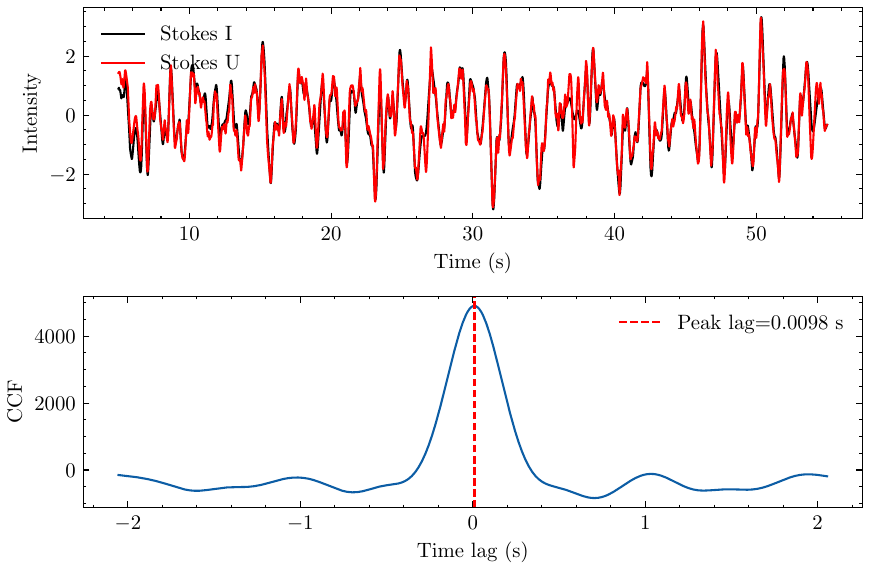}
    \includegraphics[width=0.45\textwidth]{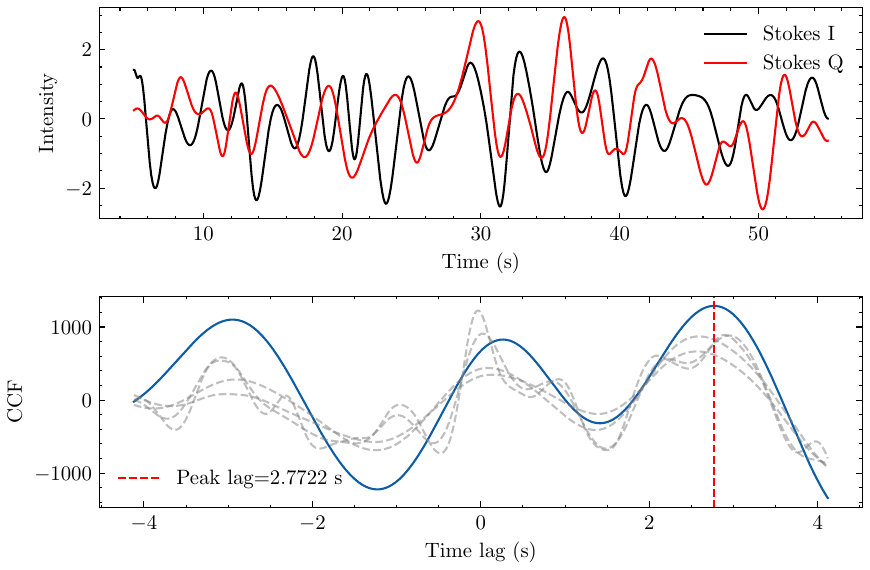}
    \caption{Left: CCF result between Stokes I and U. The peak occurs at a lag of 0.0098 s, which is below the temporal resolution of the observation. This indicates that no significant time delay is detected between the light curves in Stokes I and U. Right: cross-correlation of Stokes I with respect to Stokes Q. A peak at $\tau \approx 2.8$ s means Stokes I is delayed relative to Stokes Q by 2.8 s.} In the lower panel, gray dashed lines show the CCFs obtained by progressively removing white noise from the Stokes I and Q time series.
    \label{fig:ccf}
\end{figure*}

Based on this understanding, we performed a cross-correlation analysis on the light curves of Stokes I, Q, and U from Figure \ref{fig:Dynamic}, using a lag range of [-10$\tau_I$,10$\tau_I$]. The results are presented in Figure \ref{fig:ccf}. No significant time delay was detected between Stokes I and U. In contrast, a clear time delay of approximately 2.8 s was measured between Stokes I and Q. Using the projected distance between the core and the SW jet component on the scattering screen, this delay corresponds to a velocity of about 673 km/s. This derived velocity is consistent with values reported in previous studies \citep{2021MNRAS.504.5437L,1995GeoRL..22.3301P}.
The IPS results presented in Figure \ref{fig:Dynamic} \& Figure \ref{fig:acf}, which confirm that Stokes Q and U originate from distinct emitting regions, have critical implications for the reduction of related FAST calibration data.

The detailed analysis of the auto- and cross-correlations is warranted to interpret the scintillation properties. For weak scintillation (modulation index $m \ll 1$) and a point source ($\theta_S = 0$), the spatial scale at half-power is approximately $r_F$, while the $1/e$ scale is about 10$\%$ longer. If the source has a finite angular diameter $\theta_S$, the spatial scale increases to roughly $\theta_S \cdot 1\,\,\mathrm{AU}$, and the scintillation index decreases by a factor proportional to the square root of the ratio $\theta_S \cdot 1\,\mathrm{AU} / r_F$.

From our data, the $1/e$ time scales for Stokes $I$ and $U$ are both $\sim 0.195\,\mathrm{s}$, implying a half-power scale $\sim 10\%$ smaller, i.e., $r_F = V_{\,\mathrm{sw}} \times 0.175\,\mathrm{s}$. This yields an estimate of the solar wind velocity $V_{\,\mathrm{sw}} \sim 1100\,\mathrm{km}\,\mathrm{s}^{-1}$, which, while somewhat higher than typical values, is not implausible. The spatial scale for Stokes $Q$ is larger by a factor of $\sim 2.9$ compared to $I$ and $U$, indicating that the $Q$ emission region is more extended. Its angular diameter would then be $\theta_S \sim (189\,\mathrm{km} \times 2.9) / 1\,\mathrm{AU} \approx 1\,\mathrm{arcsec}$, consistent with the absence of resolved maps at higher frequencies. This broader spatial scale reduces the variance by about a factor of two, and hence the rms by $\sqrt{2}$; therefore, knowledge of the scintillation indices of $I$, $U$, and $Q$ is valuable for constraining the source structure. We have measured these indices from the ACFs (see Figure \ref{fig:acf}). While IPS theory predicts a lower scintillation index for more extended emission regions, we find that the measured index of Stokes $Q$ is larger than that of Stokes $I$. This behavior likely arises from a combination of instrumental polarization, beam-shape asymmetry, and calibration uncertainties, to which Stokes $Q$ is particularly sensitive as a differential quantity. Moreover, the uncorrelated nature of IPS-induced fluctuations in Stokes $Q$ and $U$ can further amplify the apparent variance of $Q$ on short timescales. Therefore, the observation $m_Q > m_I$ does not contradict the IPS interpretation but rather underscores the complexity of polarization-resolved scintillation measurements with FAST.

The cross-correlation would show a peak delay determined by the component of the solar wind velocity along the baseline direction (baseline of $1885\,\mathrm{km}$ for SW component). For our target 3C286 (RA $13^{\,\mathrm{h}}5$, Dec $+30^\circ$) observed on September 30, the Sun was near RA $12^{\,\mathrm{h}}$, Dec $0^\circ$, i.e., southwest of the source. Hence the solar wind flows from the southwest toward the northeast, passing first through the $Q$-emitting region (presumably located southwest of the core) and then through the $I$-emitting region. This geometry predicts that intensity variations in $I$ should lag those in $Q$. Using the convention that a positive lag in the cross-correlation $C(\tau) = \langle I(t) Q(t+\tau) \rangle$ corresponds to $I$ lagging $Q$, we indeed observe a prominent peak at $\tau \approx +2.8\,\mathrm{s}$, confirming that $I$ lags $Q$ by this amount. A smaller peak near zero lag may arise from $Q$ emission in the eastern jet, while another feature near $-2.8\,\mathrm{s}$ is not readily explained by simple geometry. The velocity inferred from the $+2.8\,\mathrm{s}$ lag is $1885\,\mathrm{km} / 2.8\,\mathrm{s} \approx 670\,\mathrm{km}\,\mathrm{s}^{-1}$, which differs from the $1100\,\mathrm{km}\,\mathrm{s}^{-1}$ estimated from the auto-correlation scales. This discrepancy can be reconciled if the solar wind turbulence is anisotropic, with spatial scales elongated in the radial direction. It is known that within $20R\odot$ the turbulence is highly anisotropic, and in the region of our observations the axial ratio may be $\sim 1.5$. If the radial scale is $2.5\,r_F$, the implied velocity becomes $\sim 700\,\mathrm{km}\,\mathrm{s}^{-1}$, in good agreement with the cross-correlation result. While the present data do not allow a more precise determination of the auto- and cross-correlations, the results are plausible and suggest that single-dish velocity estimates could be extended to other multi-component radio sources where the components can be separated via polarization or frequency.

According to the discussion above, two practical strategies can be employed to mitigate the impact of IPS on calibrator observations: first, conducting the observation campaign when the targeting source is at a larger solar elongation; and second,
reducing the temporal resolution or increasing the integration time of radio observations \citep{2013ApJS..204...19P,2016ApJ...824..132N}. Furthermore, power density spectrum of IPS reveals that radio waves propagating through the solar wind exhibit scintillation predominantly in the 1-3 Hz frequency range \citep{2015SoPh..290.2539M,2019SpWea..17.1114C,2021MNRAS.504.5437L,2023FrASS..1059166X}, This implies that observations conducted with a time resolution greater than $\sim$ 1 seconds would be minimally affected by IPS.

The PA in Figure \ref{fig:lightcurves} is significantly inconsistent with previous observations \citep{1994A&A...284..331O,2013ApJS..204...19P,2013ApJS..206...16P,2017ApJS..230....7P}. This deviation in PA is likely attributed to the fact that FAST's beam center was aimed to the core of 3C 286, which primarily contributes to Stokes U and I, while the two-sided jets responsible for Stokes Q were offset by 0.8 arcsec and 2.6 arcsec, respectively, from the central core. Such offsets would inevitably lead to a rapid falling response of FAST's receiver for Stokes Q, resulting in an underestimated measurement of Stokes Q and, consequently, a deviation in the derived PA \citep{2020RAA....20...64J}. Therefore, this PA result should originate from the core, where the PA is measured approximately at $\sim$ 40-50 degrees \citep{2004ChJAA...4...28A,2016ApJ...824..132N}. Other factors may also contribute to the PA deviation, such as limitations of calibrating with a single source. A full correction of beam shape is not feasible with the current observations. We therefore report the observed PA while explicitly noting these possible causes and limitations, and in future FAST observations we plan to include additional calibrator sources and optimize the observing strategy for 3C 286 to further investigate and minimize these effects.


\section{Conclusion}\label{sec:con}
In this work, we examined all available FAST observations of the calibrator 3C 286 conducted between 2019 and 2023, finding prominent IPS signals present in the datasets, and observed distinct scintillation patterns across all four Stokes parameters in both dynamic spectra and light curves. Specifically, the IPS in Stokes I and U appeared simultaneously with identical drifting rates in dynamic spectra, while those in Stokes Q showed no direct correlation with signals in I and U. This phenomenon primarily results from the kiloparsec-scale morphology of 3C 286, where its radio core dominates the polarized emission in Stokes U and its southwestern jet contributes to Stokes Q. Both emitting regions have angular size of approximately 1 arcsec, consistent with the characteristic scale of radio sources that may induce IPS. Their 2.6 arcsec separation projects to a physical distance of $\sim$ 1885 km at 1 AU, significantly larger than the Fresnel scale ($\sim$ 189 km), indicating they can be treated as independent IPS point sources. Considering FAST's beam FOV of 2.9 arcmin, the telescope simultaneously can consequently detect IPS signals from both components, thereby explaining the observed differences in scintillation patterns between Stokes Q and Stokes I, U and confirming this interpretation.Moreover, the time delay between the IPS in Stokes I and Q provides a direct measure of the solar wind transit velocity during the observation.

\section{Acknowledgements}
We are gratitude to the referee for the suggestions that improve the manuscript, and Paul Demorest and Rick Perley for their help. This work is supported by the NSFC (12133007), the National Key Research and Development Program of China (Grants No. 2021YFA0718503 and 2023YFA1607901), the Youth Program of Natural Science Foundation of Hubei Province (2024AFB386), and China Postdoctoral Science Foundation (GZC20241282, 2025M773195).

\bibliography{sample7}{}
\bibliographystyle{aasjournalv7}


\end{document}